\newcommand{\beq}{\begin{equation}}
\newcommand{\eeq}{\end{equation}}
\newcommand{\bea}{\begin{eqnarray}}
\newcommand{\eea}{\end{eqnarray}}

\newcommand{\gsim}{\lower.7ex\hbox{$\;\stackrel{\textstyle>}{\sim}\;$}}
\newcommand{\lsim}{\lower.7ex\hbox{$\;\stackrel{\textstyle<}{\sim}\;$}}

\newcommand{\mrm}{\mathrm}

\documentclass[aps,prev,twocolumn,preprintnumbers,floatfix,nofootinbib]{revtex4-1}
\pdfoutput=1
\usepackage{graphicx}
\usepackage{epstopdf}
\usepackage{mathrsfs}
\usepackage{amssymb}
\usepackage{verbatim}
\usepackage{color}
\usepackage{multirow}

\def\stacksymbols #1#2#3#4{\def\theguybelow{#2}
    \def\vp{\lower#3pt}
    \def\sp{\baselineskip0pt\lineskip#4pt}
    \mathrel{\mathpalette\intermediary#1}}

\def\intermediary#1#2{\vp\vbox{\sp
     \everycr={}\tabskip0pt
     \halign{$\mathsurround0pt#1\hfil##\hfil$\crcr#2\crcr
              \theguybelow\crcr}}}

\def\be{\begin{equation}}
\def\ee{\end{equation}}
\def\bea{\begin{eqnarray}}
\def\eea{\end{eqnarray}}

\def\sp{\;\;\;,\;\;\;}

\def\mrm{\mathrm}

\def\lsim{\raise0.3ex\hbox{$\;<$\kern-0.75em\raise-1.1ex\hbox{$\sim\;$}}}
\def\gsim{\raise0.3ex\hbox{$\;>$\kern-0.75em\raise-1.1ex\hbox{$\sim\;$}}}

\def\inbar{\,\vrule height1.5ex width.4pt depth0pt}

\def\IC{\relax\hbox{$\inbar\kern-.3em{\rm C}$}}
\def\IQ{\relax\hbox{$\inbar\kern-.3em{\rm Q}$}}
\def\IR{\relax{\rm I\kern-.18em R}}
 \font\cmss=cmss10 \font\cmsss=cmss10 at 7pt
\def\IZ{\relax\ifmmode\mathchoice
 {\hbox{\cmss Z\kern-.4em Z}}{\hbox{\cmss Z\kern-.4em Z}}
 {\lower.9pt\hbox{\cmsss Z\kern-.4em Z}}
 {\lower1.2pt\hbox{\cmsss Z\kern-.4em Z}}\else{\cmss Z\kern-.4em Z}\fi}

\def\comment#1{}
\def\to{\rightarrow}

\def\u1x{U(1)_X}
\newcommand{\nc}{\newcommand}
\nc{\LL}{L}
\nc{\vv}{\tilde{v}}
\nc{\ccdot}{\!\cdot\!}
\nc{\gsm}{G_{SM}}
\nc{\vfive}{\mathbf{5}\oplus\mathbf{\overline{5}}}
\nc{\vten}{\mathbf{10}\oplus\mathbf{\overline{10}}}
\nc{\zhol}{Z^{\rm hol}}
\nc{\xfb}{\,{\rm fb}}

\begin{document}

%
%

\preprint{CPHT-RR010.032017}
\preprint{LPT--Orsay 17-12}
\preprint{UMN--TH--3621/17}
\preprint{FTPI--MINN--17/05}

\vspace*{1mm}

\title{The Case for an EeV Gravitino}

\author{Emilian Dudas$^{a}$}
\email{Emilian.Dudas@cpht.polytechnique.fr}
\author{Yann Mambrini$^{b}$}
\email{yann.mambrini@th.u-psud.fr}
\author{Keith A. Olive$^{c}$}
\email{olive@physics.umn.edu}

\vspace{0.1cm}
\affiliation{
${}^a$ CPhT, Ecole Polytechnique, 91128 Palaiseau Cedex, France }
\affiliation{
${}^b$ Laboratoire de Physique Th\'eorique 
Universit\'e Paris-Sud, F-91405 Orsay, France.
 }
 \affiliation{
${}^c$ 
 William I.~Fine Theoretical Physics Institute, 
       School of Physics and Astronomy,
            University of Minnesota, Minneapolis, MN 55455, USA
}

\begin{abstract} 

We consider the possibility that supersymmetry is broken above the inflationary mass scale 
and that the only ``low" energy remnant of supersymmetry is the gravitino with mass of order the EeV scale.
The gravitino in this class of models becomes a candidate for the dark matter of the Universe.
To avoid the over-production of gravitinos from the decays of the next-to-lightest supersymmetric particle
we argue that the supersymmetric spectrum must lie above the inflationary mass scale 
($M_{\rm SUSY} > 10^{-5} M_{\rm P} \sim 10^{13}$ GeV). Since $m_{3/2} \simeq M_{\rm SUSY}^2/M_{\rm P}$,
we expect $m_{3/2} \gtrsim 0.2$ EeV. Cosmological constraints then predict a relatively large reheating temperature between $10^{10}$ and $10^{12}$ GeV.
\\
\begin{center}
{\it \large Dedicated to the memory of Pierre Bin\'etruy}
\end{center}

\end{abstract}

\maketitle


\maketitle


\setcounter{equation}{0}



\section{Introduction}

To date, there is no significant experimental signal for weak scale (TeV) supersymmetry at the LHC \cite{nosusy}.  
In parallel, direct detection experiments such as XENON100 \cite{XENON}, LUX \cite{LUX} or PandaX \cite{PANDAX} have set strong limits on the elastic scattering cross section
of neutralinos on nucleons exceeding common pre-run I LHC predictions \cite{mc3,mc12}. 
This may indicate one of the following:
1) low energy supersymmetry is still around the corner waiting to be discovered at a slightly 
higher energy scale \cite{mc12,beyond}; 2) part of the supersymmetric spectrum
lies at very high energy as in split supersymmetry \cite{split}; 3) essentially the entire
supersymmetric spectrum lies at very high energy as in supersplit supersymmetry \cite{supersplit}
(aka the Standard Model).  Here,  we consider the possibility that the only remnant of supersymmetry
surviving down to energies significantly below the Planck scale\footnote{We will consider the reduced Planck mass $M_{\rm P}^2 = 1/8\pi G_N \simeq 2.4\times 10^{18}$ GeV throughout the paper.} is the gravitino.

The gravitino may either be an excellent dark matter candidate \cite{pp,nos,ehnos,oss,eoss5,0404231,stef,buch,rosz,covi} or a severe cosmological problem \cite{prob,ego}.
If the gravitino is the lightest supersymmetric particle (LSP) and therefore a dark matter
candidate, there is the risk of overproduction from the decay 
of the next-to-lightest supersymmetric particle (NLSP) \cite{myy,kmy,ego}.
In fact, as we discuss below, the upper limit on the NLSP mass of several TeV
allows us to place an upper limit of  $\simeq 4$ TeV on the gravitino mass.
However, if the sparticle spectrum lies above the inflationary mass scale,
and none of the superpartners are ever produced after inflationary reheating, 
the gravitino may once again become a dark matter 
candidate with a mass of order the EeV scale.  Note that such a spectrum implies that supersymmetry is nonlinearly realized \cite{nilpotent}.  

The letter is organized as follows. In the next section, we discuss limits on the gravitino mass in
typical supersymmetric models. We discuss both the limits from big bang nucleosynthesis \cite{bbn} 
and from NLSP decay. In Section III, we consider a high scale supersymmetric model where
only the gravitino lies below the inflationary mass scale.  We derive a new lower limit to the gravitino mass 
in this case.  Assuming that the gravitino is the dark matter, we consider general 
consequences for inflationary models, particularly aspects of reheating. 
Prospects and conclusions are summarized in Section IV.


\section{Upper limits to the gravitino mass in typical SUSY scenarios }
\label{sec:model}

The physics behind the limits on the gravitino mass can be very different depending
on the specific mass range under consideration. With the exception of the cases of light
(MeV, keV, or sub-keV) masses, typical gravitino masses discussed in the literature
are in the 10-1000 GeV range similar to the masses expected for MSSM superpartners
if the SUSY scale is related to the hierarchy problem. 
However, it is well known that a gravitino with O(100) GeV mass is potentially problematic \cite{prob,ego}.
On the one hand, if it is not the LSP, it will decay to lighter sparticles, and if it is the LSP,
the NLSP would decay to the gravitino.  In
either case, the lifetime may easily fall within the range of $100 - 10^8$ s and be subject to constraints 
from BBN \cite{bbn,kkm,ceflos,ceflos2,ps,SFT,0804}. 
For example, the decay rate of a neutralino NLSP to a gravitino and photon is given by
\cite{eoss5,0404231,0804}
\beq
\Gamma_{\rm decay} \simeq \frac{C^2}{16 \pi} \frac{m_\chi^5}{m_{3/2}^2 M_P^2} 
\label{gammadecay}
\eeq
where $C$ depends on the neutralino diagonalization matrix and we have ignored
phase space factors (and other factors of O(1)). 
In the case of a gravitino LSP, there are typically strong constraints on the SUSY parameter space
forcing one into regions where the NLSP is the tau slepton \cite{ceflos2,ps}.

The BBN constraints begin to be relaxed when the lifetime of the NLSP becomes less than
O(100) s \cite{kkm,ceflos}, and for a neutralino NLSP, we can use Eq.(\ref{gammadecay})
to obtain a relation between the neutralino and gravitino masses,
\beq
\tau_\chi \lesssim 100~\rm{s.} ~~\Rightarrow ~~m_\chi > 300 ~ {\rm GeV} \left( \frac{m_{3/2}}{{\rm GeV}}\right)^{2/5} 
\label{llimit}
\eeq
for $C\sim 1$.  Thus avoiding the limits from BBN will require a rather heavy
SUSY spectrum for TeV scale (and above) gravitino masses. We note that the 
relaxation of the BBN bound at 100 s requires satisfying the upper bound on
the density of decaying particles of roughly \cite{kkm}, $m_\chi n_\chi/n_\gamma \lesssim 7 \times 10^{-9}$ GeV.
If we exceed this density, we must use the more strict BBN bound of $\tau_\chi \lesssim 0.1 s$. In this case, the lower limit on $m_\chi$ in Eq.(\ref{llimit}) is increased by a factor of $\sim 4$.

In addition to the BBN constraints, there is an additional constraint coming from the 
relic density of the NLSP whose decay contributes to the relic density of gravitinos \cite{myy,kmy,ego}. 
The gravitino relic density from NLSP decays can be written simply as 
\beq
\Omega_{3/2} h^2 = \frac{m_{3/2}}{m_\chi} \Omega_\chi h^2 
\eeq
and thus the NLSP relic density is limited by 
\beq
\Omega_\chi h^2 \lesssim 0.12  \frac{m_\chi}{m_{3/2}}
\eeq
where 0.12 is the approximate upper limit on the cold dark matter density from PLANCK experiment \cite{planck}.
As long as $m_\chi$ is not much greater than $m_{3/2}$, the NLSP density is constrained to be 
near the cold dark matter density. Even in the event that $m_\chi \gg m_{3/2}$,
the relic density of the NSLP is still constrained by the BBN unless its lifetime is very short ($< 0.1$ s)  as noted above. 

Thus as we attempt to increase the mass of a gravitino LSP, 
we are forced to higher NLSP masses to insure both a relatively short lifetime and low relic density.
For example, for $m_{3/2} = 2$ TeV, we must require 
$m_\chi \gtrsim 6$ TeV (20 TeV) to obtain $\tau_\chi < 100$ s ($ < 0.1 s$).
Generally, it is very difficult to obtain an acceptable
neutralino 
relic density when the neutralino masses surpass the TeV scale \cite{mc12,beyond}.
In particular, the neutralino relic density in the TeV regime must be regulated by
either some strong resonant process or co-annihilation. Indeed, the strongest such process 
involves the co-annihilation with the gluino \cite{glu,deSimone:2014pda,ELO,eelo}.
Pushing the mass scales to their limit (when the neutralino and gluino masses are degenerate),
an upper limit to the neutralino mass of roughly 8 TeV was found \cite{deSimone:2014pda,ELO,eelo}.
This translates (using Eq. \ref{llimit}) to an upper bound on the gravitino mass of roughly
$m_{3/2} < 4$ TeV.

\section{High scale SUSY breaking and Inflation - EeV scale gravitinos}

\subsection{High scale SUSY}

In order to go beyond the derived upper limit on the gravitino mass of 4 TeV,
we must make a more substantial departure from the common paradigm of weak 
scale supersymmetry.  In this section, we consider the possibility for a higher
gravitino masses along with a very high SUSY breaking scale, leaving only the gravitino
surviving at low energies as a dark matter candidate.

As we demonstrated in the previous section, a gravitino mass in excess of 4 TeV, 
would require a SUSY spectrum in excess of 8 TeV in order to obtain 
NLSP lifetimes short enough to be compatible with constraints from BBN.
However, even in the limit of degenerate neutralinos and gluinos, 
strong co-annihilations are insufficient to lower the NLSP relic density to acceptable levels.
Further increasing the SUSY mass scale, weakens the interaction strengths, 
lowering the annihilation (and co-annihilation) cross sections, leading to an overabundance.
Without resorting to some unknown form of dilution, one possibility for larger 
gravitino masses is to move the SUSY matter spectrum to such high scales,
so that SUSY particles were never part of the thermal bath after inflation.

To completely remove the supersymmetric particle spectrum from the thermal history,
we must assume that the SUSY mass spectrum is larger than both the 
inflationary reheating temperature, $T_R$, and the inflaton mass, $m_\phi$,
so as to prevent SUSY particles from being produced by either thermal 
processes during reheating or by the decay of the inflaton.
Here, we will not tie ourselves to a particular inflationary model,
but note that in many models considered, the inflaton mass
is set by amplitude of density perturbations seen in the microwave background,
and yields a value of roughly $3 \times 10^{13}$ GeV. When we need to refer to a specific example,
we consider a no-scale supergravity model of inflation \cite{eno} which leads to 
Starobinsky-like inflation \cite{Staro}. 

If we denote as $F$ the order parameter for supersymmetry breaking, then typical soft SUSY masses  will be proportional to $F$,
\beq
M_{SUSY} = \frac{F}{\Lambda_{mess}}
\eeq
where $\Lambda_{mess}$ is the mass scale associated with the mediators of supersymmetry breaking\footnote{These messengers could in principle also play a role in restoring unification at high scale.}.
We expect $\Lambda_{mess} \ge M_{SUSY}$. 
Thus $M_{SUSY} > m_\phi$ translates to $F > m_\phi^2$.
The gravitino mass is also determined by $F$ \cite{dz},
\beq
m_{3/2} = \frac{F}{\sqrt{3} M_P}
\eeq
And hence we have a {\em lower} bound on the gravitino mass given by
\beq
m_{3/2} > \frac{m_\phi^2}{\sqrt{3}M_P}\simeq 0.2~ {\rm EeV}
\label{Eq:mgmin}
\eeq
Thus we have a gravitino mass gap between 4 TeV and 0.2 EeV which remains 
cosmologically problematic.

\subsection{Gravitino Production}
Clearly the LHC  bounds can be satisfied if the  sparticle mass spectrum lies
above a few TeV.  The  direct detection limits can also be satisfied  as the spectrum approaches
its upper limit \cite{beyond}.
It is also possible that the dark matter lies beyond the MSSM and has weaker couplings to matter, e.g. through a t-channel exchange of a massive Z' or Higgs as shown in \cite{Arcadi:2017kky} or invoking a pseudoscalar or pure axial  mediator to velocity suppress $\sigma^{scat}_N$ \cite{Zportal, Zpportal}. 
Furthermore, if the dark matter couples too weakly with the standard model, it will never reach thermal equilibrium as its production rate is $\frac{dn}{dt} = n_\gamma^2 \langle \sigma v\rangle$. The particle is frozen {\em in} during the process of thermalization. The weak coupling of the dark sector with the standard model can be due to either an effectively small coupling (of the order of $10^{-10}$ ) \cite{fimp} or because the mass of the mediator between the two sectors is very large, as in the case of Non-Equilibrium Thermal Dark Matter (NETDM) models \cite{Mambrini:2013iaa}. 

By increasing the SUSY mass scale, we have also removed most of the standard 
gravitino production mechanisms. Namely both NSLP decay, and the thermal production
from standard model annihilations such as gluon, gluon $\to$ gluino, gravitino are no longer kinematically allowed.
The rate for the latter is well known \cite{bbb,egnop} and scales as $\Gamma \sim T^3 M_{SUSY}^2/M_P^2 m_{3/2}^2$, where we have assumed predominantly goldstino production in the limit $m_{3/2} \ll M_{SUSY}$.
In this case, the gravitino abundance is approximately $n_{3/2}/n_\gamma \sim \Gamma/H \sim T M_{SUSY}^2/M_P m_{3/2}^2$, where we have simply taken the Hubble parameter as $T^2/M_P$.

In the limit that the SUSY mass scale is above the inflationary scale, there remains, however, (at least) two sources of gravitino production.  Inflaton decay to gravitinos \cite{egnop,egno4}, and thermal production 
of two gravitinos from the thermal bath (gluon, gluon $\to$ gravitino, gravitino) \cite{bcdm}
as this is only kinematically allowed channel.
A careful computation of the gravitino production rate was derived in 
\cite{bcdm}
\beq
R = n^2 \langle \sigma v \rangle \simeq 21.65 \times \frac{T^{12}}{F^4}
\label{Eq:r}
\eeq
where $n$ is the number density of incoming states and we see that the rate has a strong dependence on temperature and is even stronger than
the NETDM case \cite{Mambrini:2013iaa} where the dependence is $R(T) \propto T^{8}$.
This dependence can be easily ascertained on dimensional grounds.  Recall that
$n \propto T^3$, and for  gravitino production, we expect  $\langle \sigma v \rangle
\propto T^6/F^4$. The consequences
of such a high temperature dependence are important: we expect
that all gravitino production will occur early and rapidly in the reheating process. This differs from the
feably coupled case \cite{fimp} where the smallness of the dark matter 
coupling to the standard model bath renders the production rate slower.

From the rate $R(T)$, we can determine that $\Gamma \sim R/n \sim T^9/M_P^4 m_{3/2}^4$ (again assuming
$m_{3/2} \ll M_{SUSY}$)
leading to a gravitino abundance $n_{3/2}/n_\gamma \sim \Gamma/H \sim T^7/M_P^3 m_{3/2}^4$.
More precisely, we find,
\beq
\Omega_{3/2}h^2 
\simeq
0.11 \left( \frac{0.1 ~\mathrm{EeV}}{m_{3/2}} \right)^3
\left( \frac{T_{RH}}{2.0 \times 10^{10}~\mathrm{GeV}} \right)^7
\label{Eq:omega}
\eeq
In the absence of direct inflaton decays, a gravitino at the lower mass limit (\ref{Eq:mgmin}) would require a
reheating temperature of roughly $3 \times 10^{10}$ GeV, above 
the upper limit allowed by the relic abundance constraint ($T_R \lesssim 10^7$ GeV) in the
more common thermal scenario \cite{bbb},
thus favoring thermal leptogenesis \cite{Giudice:2003jh}.

\subsection{Consequences for inflationary models}

\begin{figure}
\centering
\includegraphics[width=0.80\columnwidth]{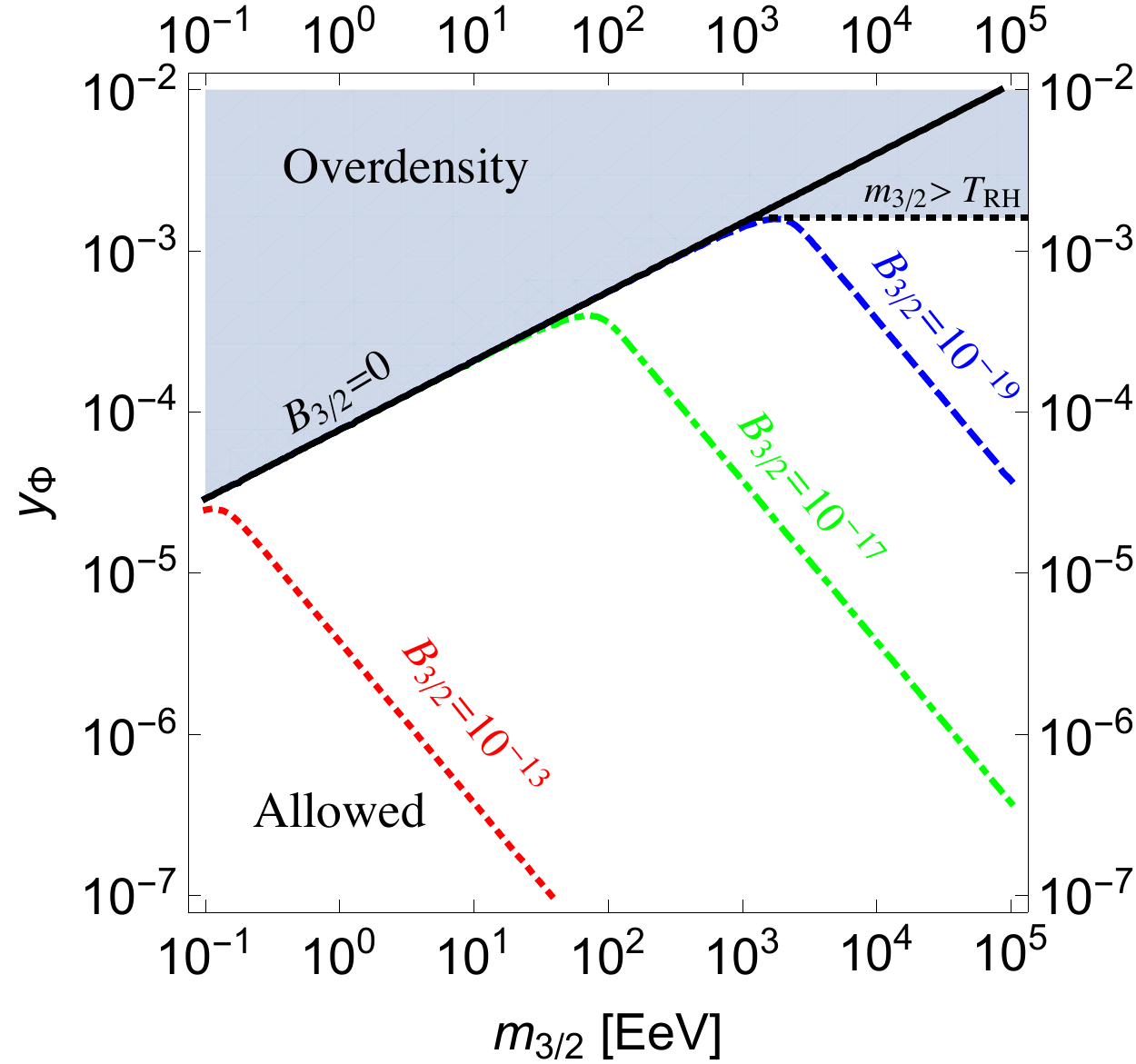}
\caption{
\footnotesize
{Region of the parameter space allowed by PLANCK constraints
\cite{planck} in the plane ($m_{3/2},y_\phi$) for different values of the branching ratio $B_{3/2}$ and 
$m_\phi=3 \times 10^{13}$ GeV (see the text for details).}
}
\label{Fig:omega}
\end{figure}

The reheating temperature appearing in Eq.(\ref{Eq:omega}) is
generated by the decay of an inflaton field $\phi$ of mass
$m_\phi$ and width $\Gamma_\phi$. 
We assume that the decay and thermalization occur instantaneously at the time $t_\phi$, 
$\Gamma_\phi t_\phi = 2 \Gamma_\phi/3H =c$, where $c \approx 1.2$ is a constant. In this case, the reheating temperature is given by \cite{ps2,egnop} 
\beq
T_{RH} = \left(\frac{10}{g_s} \right)^{1/4} \left(\frac{  2 \Gamma_\phi ~M_P}{\pi~ c} \right)^{1/2} = 0.55 \frac{y_\phi}{2\pi} \left( \frac{m_\phi~M_P}{c } \right)^{1/2}
\label{Eq:trh}
\eeq
where we have defined a standard "yukawa"-like coupling $y_\phi$ of the inflaton field to the thermal bath, 
$\Gamma_\phi = \frac{y_\phi^2}{8 \pi} m_\phi$ and $g_s$ is the effective number of light degrees
of freedom in this case set by the Standard Model, $g_s = 427/4$.
We can then re-express the relic abundance (\ref{Eq:omega})
as function of $y_\phi$:
\beq
\Omega_{3/2} h^2 \simeq 0.11 \left( \frac{0.1 ~\mrm{EeV}}{m_{3/2}} \right)^3
\left( \frac{m_\phi}{3 \times 10^{13} {\rm GeV}} \right)^{7/2} \left( \frac{y_\phi}{2.9 \times 10^{-5}} \right)^7
\label{Eq:omegay}
\eeq
where we have set $c=1.2$. 
The cosmological constraint is plotted in  Fig.(\ref{Fig:omega}) in the ($m_{3/2}$, $y_\phi$)
plane, where we show the region allowed by PLANCK \cite{planck}.
The black (solid) line represents
the PLANCK constraint $\Omega h^2 = 0.11$. One immediately sees the
linear increase in the Yukawa coupling $y_\phi$ with
increasing gravitino mass in order to counterbalance the weakening
of the effective coupling $1/F$ responsible for its production
in the thermal bath.

A  large inflaton-matter coupling  produces a high reheating temperature,
which in turn increases the gravitino abundance.
Then, as one can see from Eq.(\ref{Eq:omegay}), the solid curve in Fig. \ref{Fig:omega} is
an upper bound on $y_\phi$ to avoid an overabundant gravitino. 
In fact, one can extract an upper bound on $y_\phi$ independent of $m_{3/2}$ simply requiring $m_{3/2} < T_{RH}$, a necessary condition for the gravitino to be thermally produced. The condition $m_{3/2} < T_{RH}$ implemented
in Eq.(\ref{Eq:omegay}) with the expression (\ref{Eq:trh}) gives
\beq
y_{\phi} \lesssim 1.6 \times 10^{-3} 
\left(\frac{3 \times 10^{13} ~\mrm{GeV}}{m_\phi}  \right)^{1/2},
\label{Eq:limityphi}
\eeq
shown as the horizontal dashed line in the Figure \ref{Fig:omega}.
We can then extract the maximum reheating temperature
$T_{RH} \lesssim 1.1 \times 10^{12}~\mrm{GeV}$. Combined with the condition (\ref{Eq:mgmin}) $m_{3/2}> $ 0.2 EeV, the relic abundance constraint (\ref{Eq:omega}) gives
\beq
2.7 \times 10^{10}~\mrm{GeV} \lesssim 
T_{RH} \lesssim 1.1 \times 10^{12} ~\mrm{GeV}
\label{Eq:trhmax}
\eeq
which is a strong prediction of our model.

\subsection{Gravitino production by inflaton decay}

It is  also possible to produce gravitinos through the direct decay of the inflaton.
For example, in no-scale supergravity models of inflation, the decay of the inflaton to gravitinos
is highly suppressed. In simple models,  there is no coupling at the tree-level \cite{ekoty}.
However, it is possible to couple the inflaton to moduli without spoiling the
inflationary potential \cite{egno4,egnop}. We can parameterize the decay to 
a pair of gravitinos as
$\Gamma_{3/2} = m_\phi \frac{y_{3/2}^2}{72 \pi}$  .

The branching ratio of decays to gravitinos is then
\beq
B_{3/2} = \Gamma_{3/2}/\Gamma_\phi= \frac{|y_{3/2}|^2}{9 y_\phi^2} .
\eeq
Using the result from \cite{egnop} for the gravitino abundance produced by inflaton decay
at the epoch of reheating, we get
\beq 
\frac{n_{3/2}}{n_\gamma}  \approx 3.6 B_{3/2} \frac{(\Gamma_\phi M_{\rm P})^{1/2}}{m_\phi} \approx 0.7 B_{3/2} y_\phi \left(\frac{M_{\rm P}}{m_\phi}\right)^{1/2}
\eeq
corresponding to
\bea
\Omega_{3/2}^{decay}h^2 = 0.11
&&\left(\frac{B_{3/2}}{1.3\times 10^{-13}}\right) \left(\frac{y_\phi}{2.9\times 10^{-5} } \right) 
\\
&&
\times
\left(\frac{m_{3/2}}{0.1~\mrm{EeV}} \right)  \left(\frac{3 \times 10^{13} ~\mrm{GeV}}{m_\phi} \right)^{1/2}.
\nonumber
\eea
today.

The condition (\ref{Eq:mgmin}) is then translated into
\beq
B_{3/2} y_\phi= \frac{|y_{3/2}|^2}{9 |y_\phi|} \lesssim 1.9 \times 10^{-18} \left(\frac{0.1~\mrm{EeV}}{m_{3/2}} \right)
\eeq
for  $m_\phi = 3 \times 10^{13}$ GeV.
Contrary to the case of thermal gravitino production, our limit to the
coupling $y_\phi$ is strengthened as $m_{3/2}$ is increased 
when gravitino production occurs through inflaton decay.
Since the density through the decay of the inflaton is proportional to $n_{\phi} B_{3/2} m_{3/2}$, where $m_\phi n_\phi$ is the inflaton energy density,  the limit on the coupling
is improved when either the branching ratio or the gravitino mass is increased. 

This result is also shown in Fig.(\ref{Fig:omega}) where we clearly see the changing in the slope for larger value of 
$B_{3/2} > 10^{-19} $ where the direct production from inflaton decay may dominate over the thermal production. We note that the constraints obtained on the inflaton coupling to gravitinos are strong. We recall, however, that in no-scale models of inflation \cite{ekoty,egno4,egnop}
and in classes of inflationary models with so-called stabilized field \cite{yanagida,terada}, this coupling is naturally very small.
Finally, we point out that in the case of the direct production of the gravitino through inflaton decay, both the $\pm 3/2$
and the $\pm 1/2$ components of the gravitino populate the Universe, whereas in the case of thermal production (Eq.\ref{Eq:omega}) only the longitudinal goldstino component  contributes to the relic abundance.

\section*{Perspectives and Conclusions}

\noindent
In many ways, it seems quite natural that a particle with only
gravitational interactions should  make up the dark matter of the Universe.
We have seen that in the generic context with gravitino dark matter where
the supersymmetric particle spectrum thermalizes with standard model
bath, an upper limit to the mass of the gravitino of
$\simeq 4$ TeV is obtained. However, if one makes the
minimal hypothesis that the supersymmetric spectrum 
lies above the inflaton mass, a new 
cosmologically allowed window opens for gravitino mass 
above 0.2 EeV. Indeed, despite the weakness of its 
coupling, the gravitino can be produced
directly from the thermal bath by the exchange of virtual
heavy superpartners (or equivalently by higher dimensional
operators). It can also be produced directly from the inflaton decay. 
In order to obtain gravitino dark matter from the thermal bath, we predict 
a relatively large reheating temperature $\gtrsim 10^{10}$
GeV, compatible with the thermal leptogenesis scenario. 
If stable,  this gravitino is virtually undetectable as it 
is the only  $R$-parity odd state ever present in the Universe after inflation.
If unstable through an  $R$-parity
violating coupling, the decay of the gravitino would produce EeV--like monochromatic
photons or neutrinos, which are not yet observable by present experiments.

\noindent {\bf Acknowledgements. }  The authors acknowledge Karim Benakli and Yifan Chen for useful discussions. This  work was supported by the France-US PICS no. 06482.
 Y.M.  acknowledges partial support from the European Union FP7 ITN INVISIBLES (Marie
Curie Actions, PITN- GA-2011- 289442) and  the ERC advanced grants  
 Higgs@LHC. E.D. acknowledges partial support from the ANR Black-dS-String. The work of K.A.O. was supported in part
by DOE grant DE--SC0011842 at the University of Minnesota.

\end{document}